\documentclass[preprint,12pt]{elsarticle}
\usepackage{graphicx}
\usepackage{enumerate}
\usepackage{amsmath}
 \usepackage{framed}
  \usepackage{setspace}
  \usepackage{amssymb}
  \usepackage[justification=centering]{caption}
\usepackage{algorithm}
\usepackage{algpseudocode}
\usepackage{comment}
\usepackage{bm}

\newtheorem{theorem}{Theorem} 
\newtheorem{definition}{Definition}
\newtheorem{lemma}{Lemma}

\newtheorem{corollary}{Corollary}
\makeatletter
\renewcommand{\ALG@beginalgorithmic}{\normalsize}
\makeatother

\usepackage{float}
\usepackage{lipsum}



\begin{document}

\begin{frontmatter}
\title{A quantum related-key attack based on Bernstein-Vazirani algorithm}


\author{Huiqin Xie$^{1,2,3}$}
\author{Li Yang$^{1,2,3}$\corref{1}}
\cortext[1]{Corresponding author email: yangli@iie.ac.cn}
\address{1.State Key Laboratory of Information Security, Institute of Information Engineering, Chinese Academy of Sciences, Beijing 100093, China\\
2.Data Assurance and Communication Security Research Center,Chinese Academy of Sciences, Beijing {\rm 100093}, China\\
3.School of Cyber Security, University of Chinese Academy of Sciences, Beijing {\rm  100049}, China}

\begin{abstract}
Due to the powerful computing capability of quantum computers, cryptographic researchers have applied quantum algorithms to cryptanalysis and obtained many interesting results in recent years. In this paper, we study related-key attack in the quantum setting, and proposed a specific related-key attack which can recover the key of block ciphers efficiently, as long as the attacked block ciphers satisfy certain conditions. The attack algorithm employs Bernstein-Vazirani algorithm as a subroutine and requires the attacker to query the encryption oracle with quantum superpositions. Afterwards, we rigorously demonstrate the validity of the attack and analyze its complexity. Our work shows that related-key attack is quite powerful when combined with quantum algorithms, and provides some guidance for the design of block ciphers that are secure against quantum adversaries.

\end{abstract}


\end{frontmatter}


\section{Introduction}
Shor' algorithm \cite{Sho94} indicates that once scalable quantum computers are available, many widely used asymmetric cryptosystems, such as RSA, will be broken. This has sparked a upsurge of research on post-quantum cryptography, which studies classical systems that are secure against quantum adversaries. In response to the threat of quantum computing, NIST has initiated the process of standardizing post-quantum public-key algorithms \cite{NIST}.

On the other hand, although less attention is paid than public-key cryptography, symmetric cryptosystems are also suffering the threat from quantum attacks. For example, due to Grover's algorithm \cite{Gro96} general exhaustive search attacks can obtain a quadratic speedup. More strikingly, some symmetric systems that have been proved to be secure against classical adversaries have been broken by polynomial-time quantum algorithms. Kuwakado and Morii made use of Simon's algorithm \cite{Sim97} to distinguish the three-round Feistel construction \cite{KM10} and recover the key in Even-Mansour cipher \cite{KM12}. Santoli \textit{et al}. \cite{SS17} and Kaplan \textit{et al}. \cite{KLLNP16} subsequently extended their results independently and applied Simon's algorithm to other symmetric primitives. All these attacks are executed in the model of quantum chosen-plaintext attack \cite{DOM11,BZ13,GHS16}, where the attacker can query the encryption oracle with superpositions.

When quantum chosen-plaintext attack has been widely studied, quantum related-key attack has also started to draw attention. Classical related-key attacks were first introduced by Biham \cite{Bih94}, and has been applied to Rijndael \cite{NJS00}, KASUMI \cite{JBD96} and other schemes. In such attacks, the attacker can query the encryptions or decryptions of messages under the keys that have some known mathematical relation with the target key. Roetteler and Steinwandt first study related-key model in the quantum setting \cite{MR15}. They showed that, under the assumption that the key of the block cipher can be uniquely determined by a small amount of accessible plaintext-ciphertext pairs, a quantum attacker can efficiently extract the key by using a quantum related-key attack. Afterwards, Hosoyamada and Aoki proposed a polynomial-time quantum algorithm that recovers the key of two-round iterated Even-Mansour scheme with only two queries to the related-key oracle \cite{AK17}. These two results show that related-key attack is powerful for quantum attackers.

In this paper, we further study the applications of quantum related-key attack to block ciphers. Based on Bernstein-Vazirani (BV) algorithm \cite{BV97}, we propose a quantum attack for recovering the key of general block ciphers. We prove that, if not requiring the time complexity to be polynomial, our attack can find out the key of an arbitrary unrestricted block cipher. Afterwards, we give two specific conditions, and demonstrate that, as long as the block cipher satisfies one of them, then our attack can effectively extract the secret key in polynomial time. Like the attack model of \cite{MR15}, we allow the attacker to query the encryption oracle with superpositions of keys. This makes the attack less practical because the ability to query with superpositions of keys is a strong requirement even for quantum adversaries. However, from the perspective of constructing ciphers, our results helps to establish criterions that a secure block cipher should meet in the post-quantum world.

\section{Preliminary}
Throughout this paper, we let $\mathbb{F}_2=\{0,1\}$, representing the finite field with characteristic 2. $E$ denotes an arbitrary block cipher with blocksize $n$ and key length $k$. When fix a secret key $s\in\mathbb{F}_2^k$, $E_s$ is a permutation from $\mathbb{F}_2^n$ to $\mathbb{F}_2^n$. We assume that $E$ can be efficiently implemented by a quantum circuit. That is, there exists a polynomial-time quantum circuit that takes as input a secret key along with a plaintext and output the corresponding ciphertext. The quantum circuit implements the following unitary operator:
$$
U_{E}:\sum_{m,x,y}|x\rangle|m\rangle|y\rangle\longrightarrow\sum_{m,x,y}|x\rangle|m\rangle|y\oplus E_{x}(m)\rangle.
$$
For the block ciphers used in practice, this assumption holds undoubtedly. Since the quantum circuit of $U_E$ does not involve the secret key $s$, the attacker can perform the unitary operator $U_E$ by himself.

Because the unitary quantum gates $\{H, CNOT, Phase, \frac{\pi}{8}\}$ form a universal gate set \cite{NC00}, we can assume that the quantum circuit implementing $U_E$ is composed of gates in this set. Here, $H$ is the Hadamard gate, $CNOT$ is the controlled-NOT gate, $Phase$ is the phase gate and $\frac{\pi}{8}$ is the $\frac{\pi}{8}$ gate (Fig.1). Let $|E|_Q$ be the number of universal gates in the quantum circuit implementing $E$. $|E|_Q$ is a polynomial of $k$ and $n$. The attacker can integrate $U_E$ into his circuits as in Fig.2
\begin{figure}[h]
  \centering
  \includegraphics[width=10cm]{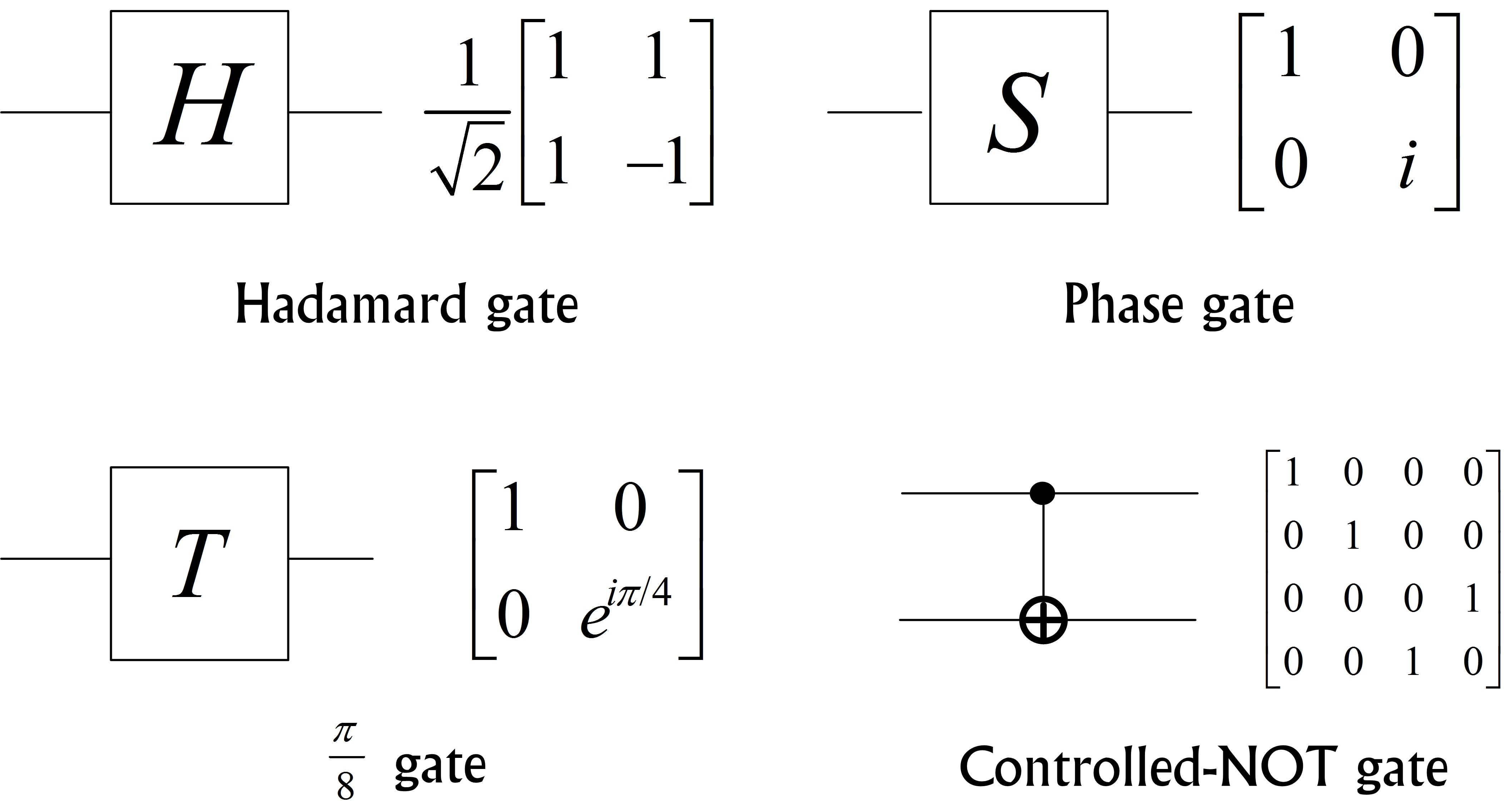}\\
  \caption{Universal gate set}
\end{figure}

\begin{figure}[h]
  \centering
  \includegraphics[width=8cm]{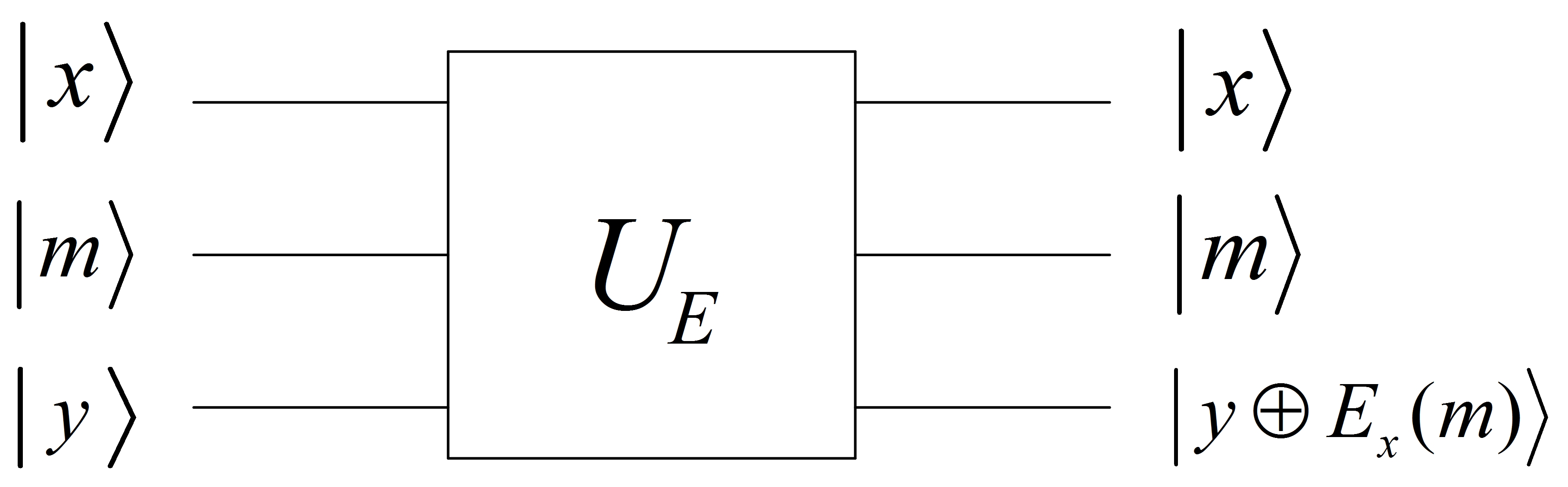}\\
  \caption{Quantum gate $U_E$\qquad\quad}
\end{figure}
\subsection{Related-key attack}
We first recall the related-key attack model proposed in \cite{RM87}, where the key relation is restricted to bit-flips. In this model, after a secret key $s\in\mathbb{F}_2^k$ is determined, the attacker can query following two oracles:
\vskip 0.2cm

$\mathcal{E}$: On input a plaintext $m\in\mathbb{F}_2^n$ and a bitmask $x\in\mathbb{F}_2^k$, $\mathcal{E}$ returns the encryption $E_{s\oplus x}(m)$.
\vskip 0.2cm

$\mathcal{D}$: On input a ciphertext $c\in\mathbb{F}_2^n$ and a bitmask $x\in\mathbb{F}_2^k$, $\mathcal{D}$ returns the decryption $E_{s\oplus x}^{-1}(c)$.
\vskip 0.2cm

\noindent
After querying these oracles, the attacker needs to output a vector $s'\in\mathbb{F}_2^k$ as a guess of $s$. He succeeds if and only if $s'=s$.

The attacks presented in this paper do not require the access to the decryption oracle $\mathcal{D}$, but the attacker is allowed to query the encryption oracle $\mathcal{E}$ with superpositions of keys. That is, the attacker can query the quantum oracle $\mathcal{O}_{\mathcal{E}}$ which operates as follows:
$$
\mathcal{O}_{\mathcal{E}}:\sum_{x,m,y}|x\rangle|m\rangle|y\rangle\longrightarrow\sum_{x,m,y}|x\rangle|m\rangle|y\oplus E_{s\oplus x}(m)\rangle.
$$
The attacker can integrate the oracle $\mathcal{O}_{\mathcal{E}}$ into his circuits as in Fig.3. Furthermore, we allow the attacker to query the oracle that returns solely a bit of the cipher with superpositions of keys. That is, supposing $E_{s\oplus x}=(E_{s\oplus x,1},E_{s\oplus x,2},\cdots,E_{s\oplus x,n})$, for each $j=1,2,\cdots,n$, the attacker can query the quantum oracle
$$
\mathcal{O}_{\mathcal{E}_j}:\sum_{x,m,y}|x\rangle|m\rangle|y\rangle\longrightarrow\sum_{x,m,y}|x\rangle|m\rangle|y\oplus E_{s\oplus x,j}(m)\rangle.
$$
\begin{figure}[H]
  \centering
  \includegraphics[width=8cm]{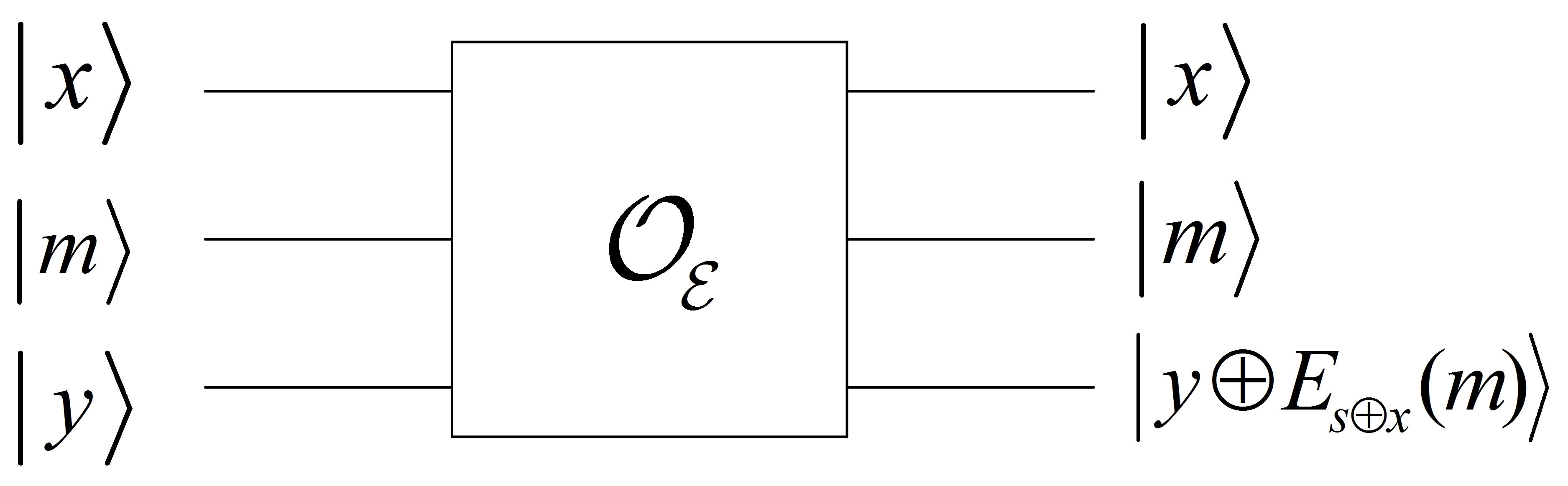}\\
  \caption{Quantum gate $\mathcal{O}_{\mathcal{E}}$\qquad\qquad\qquad}
\end{figure}
The scenario where quantum attackers can query cryptographic primitives with quantum superpositions has been considered in a significant amount research  \cite{DOM11,BZ13,GHS16,IJB13,Unr12,Zha12,Wat09}. The access to the oracle $\mathcal{O}_{\mathcal{E}}$ implies that the attacker can query the encryption oracle equipped the target key $s$. That is, the attacker can query the following oracle:
$$
\mathcal{O}_{E_s}:\sum_{m,y}|m\rangle|y\rangle\longrightarrow\sum_{m,y}|m\rangle|y\oplus E_s(m)\rangle.
$$
To do this, he only needs to query $\mathcal{O}_{\mathcal{E}}$ with the state $\sum_{m,y}|{\bm 0}\rangle|m\rangle|y\rangle$ and discard the first register. Therefore, quantum related-key attack model can be viewed as an extension of the quantum chosen-plaintext attack model.

\subsection{Linear structure}
Let $C_{k,n}$ denote the set of maps from $\mathbb{F}_2^k$ to $\mathbb{F}_2^n$. The notion of linear structure is defined as following:
\begin{definition}[\cite{OK94}]$F\in C_{k,n}$. A vector $a\in\mathbb{F}_2^k$ is said to be a linear structure of $F$ if there exist $\alpha\in\mathbb{F}_2^n$ such that
$$
F(x)\oplus F(x\oplus a)=\alpha,\,\,\,\,\forall x\in\mathbb{F}_2^k.
$$
\end{definition}

Let $U_F$ denote the set of all linear structures of $F$, and $U_F^{\alpha}:=\{a\in\mathbb{F}_2^k|F(x)\oplus F(x\oplus a)=\alpha,\,\forall x\in\mathbb{F}_2^k\}$, then $U_F=\bigcup_{\alpha}U_{F}^{\alpha}$.

\begin{definition}$F\in C_{k,n}$. A vector $a\in\mathbb{F}_2^k$ is said to be a $\sigma$-close linear structure of $F$ if there exist $\alpha\in\mathbb{F}_2^n$ such that
$$
\frac{|\{x\in\mathbb{F}_2^k|F(x)\oplus F(x\oplus a)=\alpha\}|}{2^k}> 1-\sigma.
$$
\end{definition}

Suppose $F=(F_1,F_2,\cdots,F_n)$, then obviously, $a$ is a linear structure of $F$ if and only if it is a linear structure of $F_j$ for each $j=1,2,\cdots,n$. To find a linear structure of $F$, we only need to find linear structures of every $F_j$ first, and then select a common linear structure. Therefore, in order to find linear structures of functions in $C_{k,n}$ for a general parameter $n$, we only need to focus on the case of $n=1$.

Linear structures of the functions in $C_{k,1}$ are determined by their Walsh spectrum, which is defined as following:
\begin{definition}Suppose $f:\mathbb{F}_2^k\rightarrow\mathbb{F}_2$ is a function in $C_{k,1}$. The Walsh spectrum of $f$ is defined as
\begin{align*}
S_f:\mathbb{F}_2^k&\longrightarrow\mathbb{F}_2\\
\omega&\longrightarrow S_f(\omega)=\frac{1}{2^k}\sum_{x\in \mathbb{F}_2^k}(-1)^{f(x)+\omega\cdot x}.
\end{align*}
which is also a function in $C_{k,1}$.
\end{definition}

Let $U_f$ be the set of the linear structures of $f$, and $U_f^i:=\{a\in\mathbb{F}_2^k|f(x)\oplus f(x\oplus a)=i,\,\forall x\in\mathbb{F}_2^k\}$ for $i=0,1$. We have $U_f=U_f^0\cup U_f^1$. Following lemma shows how to determine the linear structures by Walsh spectrum:
\begin{lemma}[\cite{Dub01}]For any $f\in C_{k,1}$, let $N_f:=\{\omega\in\mathbb{F}_2^k|S_f(\omega)\neq0\}$. Then for $\forall i\in\{0,1\}$, it holds that
$$
U_f^i=\{a\in\mathbb{F}_2^k|a\cdot\omega=i,\,\forall\,\omega\in N_f\}.
$$
\end{lemma}

According to the above lemma, if one has a large enough subset $W$ of $N_f$, he can solve the linear equation group $\{x\cdot\omega=i|\omega\in W\}$ to obtain the linear structures of $f$. As discussed previously, by applying this method to find each $F_j$'s linear structures, one is expected to get the linear structures of $F$. (Here solving the linear equation group $\{x\cdot\omega=i|\omega\in W\}$ means seeking vectors $x$ such that $x\cdot\omega=i$ for $\forall\omega\in W$.)

\subsection{Bernstein-Vazirani algorithm}

Given the quantum oracle access of a function $f(x)=a\cdot x$, where $a\in\mathbb{F}_2^k$ is a secret string, BV algorithm's \cite{BV97} original goal is to find $a$. However, Li and Yang observed that, when BV algorithm is applied to a general Boolean function $f:\mathbb{F}_2^k\rightarrow\mathbb{F}_2$ in $C_{k,1}$, it will always return a vector in $N_f$ \cite{LY18}. BV algorithm is executed as following:
\begin{enumerate}[  1.]
\item Perform Hadamard operator $H^{(k+1)}$ on the initial state $|\psi_0\rangle=|0\rangle^{\otimes k}|1\rangle$ to get
$$|\psi_1\rangle=\sum_{x\in \mathbb{F}_2^k}\frac{|x\rangle}{\sqrt{2^k}}\cdot\frac{|0\rangle-|1\rangle}{\sqrt{2}}.$$
\item Query the oracle of $f$ , obtaining
\begin{align*}
|\psi_2\rangle=\sum_{x\in \mathbb{F}_2^k}\frac{(-1)^{f(x)}|x\rangle}{\sqrt{2^k}}\frac{|0\rangle-|1\rangle}{\sqrt{2}}.
\end{align*}
\item Perform the Hadamard operator $H^{(k)}$ to the first $k$ qubits and discard the $(k+1)$-$th$ qubit, producing
\begin{align*}
|\psi_3\rangle&=\sum_{y\in \mathbb{F}_2^k}(\frac{1}{2^k}\sum_{x\in \mathbb{F}_2^k}(-1)^{f(x)+y\cdot x})|y\rangle\\
&=\sum_{y\in \mathbb{F}_2^k}S_f(y)|y\rangle.
\end{align*}
By measuring $|\psi_3\rangle$ in the computational basis, one will obtain a vector $y\in\mathbb{F}_2^k$ with a probability of $S_f(y)^2$.
\end{enumerate}

When applying BV algorithm to a function $f\in C_{k,1}$, it always returns a vector in $N_f$. In light of this fact and Lemma 1, one can use BV algorithm to find linear structures of an arbitrary function in $C_{k,1}$. Executing BV algorithm needs a total of $2k+1$ Hadamard gates and one quantum query. The number of qubits required is $k+1$. The quantum circuit of BV algorithm is presented in Fig.4.
\begin{figure}[H]
  \centering
  \includegraphics[width=8cm]{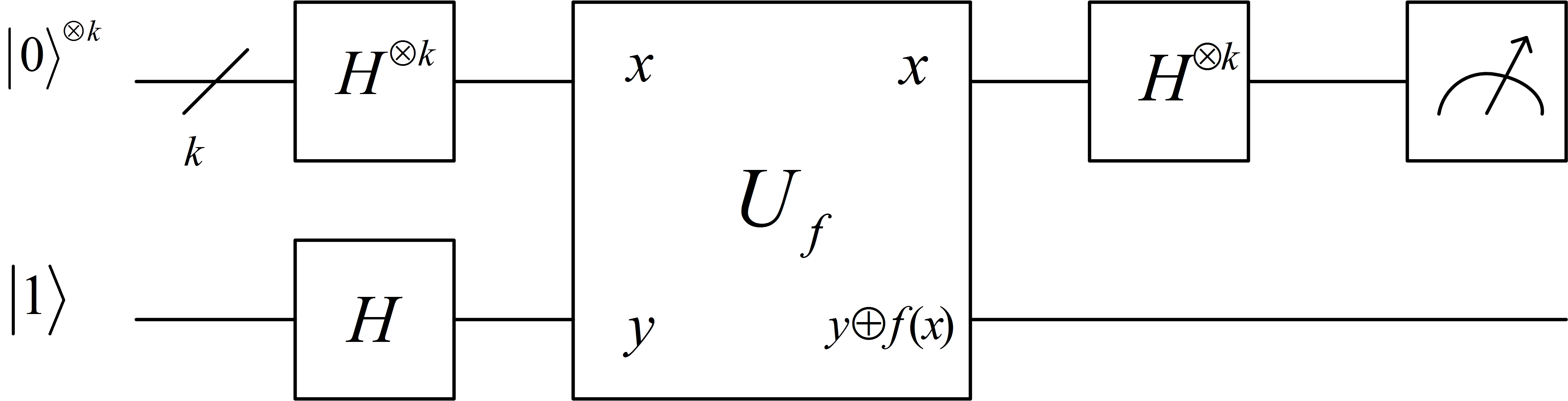}\\
  \caption{Quantum circuit of Bernstein-Vazirani algorithm}
\end{figure}

\section{Quantum algorithm for finding linear structures}

A quantum algorithm for finding nonzero linear structures of functions in $\mathcal{C}_{k,n}$ was proposed by Xie and Yang \cite{XY17}. Suppose $F=(F_1,F_2,\cdots, F_n)\in\mathcal{C}_{k,n}$. For each $j=1,2,\cdots,n$, their algorithm first calls BV algorithm to get a subset of $N_{F_j}$, then uses the subset to compute linear structures of $F_j$ according to Lemma 1. Afterwards, the algorithm selects an nonzero common linear structure of $F_1, F_2,\cdots, F_n$ and outputs it. The output vector has a high probability of being a linear structure of $F$. We make a minor modification to the algorithm in \cite{XY17} so that it outputs a set containing all linear structures of $F$, instead of only a random linear structure. The modified algorithm is as following:
\vskip 0.1cm

\begin{framed}
\noindent
Algorithm \textbf{FindStruct}
\vskip 0.2cm

\noindent
\textbf{Initialization:} $p(n)$ is a polynomial of $n$ chosen by the attacker. $F=(F_1,F_2,\cdots,F_n)\in C_{k,n}$. The quantum oracle access of each $F_j$ ($1\leq j\leq n$) is given.
\vskip 0.05cm

\noindent
\hangafter 1
\hangindent 1.35em
1. For each $j=1,2,\cdots,n$, apply BV algorithm to $F_j$ for $p(n)$ times to obtain a subset $W_j$ of $N_{F_j}$. The size of $W_j$ is $p(n)$.

\noindent
\hangafter 1
\hangindent 1.35em
2. For each $j=1,2,\cdots,n$, solve the linear equation group $\{x\cdot\omega=i_j|\omega\in W_j\}$ to obtain the solution $A^{i_j}_j$ for $i_j=0,1$, respectively. Let $A_j=A_j^0\cup A_j^1$.

\noindent
\hangafter 1
\hangindent 1.35em
3. Find the intersection $\bar{A}=A_1\cap A_2\cap\cdots\cap A_n$. For each $a\in\bar{A}$, let $\tilde{a}=(i_1,i_2,\cdots,i_n)$, where $i_1,i_2,\cdots,i_n$ are the corresponding superscripts such that $a\in A_1^{i_1}\cap A_2^{i_2}\cap\cdots\cap A_n^{i_n}$. Let $A=\{(a,\tilde{a})|a\in\bar{A}\}$ and output $A$.
\end{framed}
\vskip 0.1cm

In the above algorithm, when the attacker computes $A_j=A_j^0\cup A_j^1$ in Step 2, he actually needs to attach a tag to each vector in $A_j$. Specifically, if $a\in A_j^0$, then a tag $i_j=0$ is attached to $a$ when it is put into the set $A_j$; if $a\in A_j^1$, then a tag $i_j=1$ is attached to $a$ when it is put into the set $A_j$. Subsequently, when the attacker computes the intersection $\bar{A}$, for each $a\in A_1\cap A_2\cap\cdots\cap A_n$, he attaches the corresponding $n$ tags $i_1,i_2,\cdots,i_n$ to $a$ when puts it into $\bar{A}$. Therefore, when calculating the set $A$, the attacker can easily obtain corresponding $\tilde{a}$ of each $a\in\bar{A}$ by tracking these tags. Using these tags is for avoiding the attacker needing to compute the intersection of $n$ sets for exponential times. With these tags, the attacker only need to compute the intersection $\bar{A}=A_1\cap A_2\cap\cdots\cap A_n$ once to obtain the set $A$. If without these tags, then the attacker needs to compute the intersections $A_1^{i_1}\cap A_2^{i_2}\cap\cdots\cap A_n^{i_n}$ to obtain the linear structures in $U_F^{(i_1,\cdots,i_n)}$ for each $i_1,i_2,\cdots,i_n\in\{0,1\}$, so he needs to compute intersection for $2^n$ times.

The following three theorems demonstrate the feasibility of the algorithm \textbf{FindStruct}. Theorems 2 and 3 have been proved in \cite{XY17} and we therefore omit the proofs.

\begin{theorem}Suppose $F=(F_1,F_2,\cdots,F_n)\in C_{k,n}$ and $a$ is an arbitrary linear structure of $F$. Let $\alpha$ be the vector such that $a\in U_F^{\alpha}$. If running the algorithm \textbf{FindStruct} on $F$ returns a set $A$, then $(a,\alpha)$ must be in the set $A$.
\end{theorem}

\textbf{Proof}. Suppose $\alpha=(\alpha_1,\alpha_2,\cdots,\alpha_n)$. Since $a\in U_F^{\alpha}$, we have that $a\in U_{F_j}^{\alpha_j}$ for each $j=1,2,\cdots,n$. According to Lemma 1, for any vector $\omega\in N_{F_j}$, it holds that $a\cdot\omega=\alpha_j$. By the properties of BV algorithm, we know that the set $W_j\subseteq N_{F_j}$, so $a$ is a solution of the linear equation group $\{x\cdot\omega=\alpha_j|\omega\in W_j\}$ for each $j=1,2,\cdots,n$. Therefore, we have that $a\in\bar{A}$ and $\alpha_1,\alpha_2,\cdots,\alpha_n$ are the superscripts such that $a\in A_1^{\alpha_1}\cap A_2^{\alpha_2}\cap\cdots\cap A_n^{\alpha_n}$, which means $(a,\alpha)\in A$.

$\hfill{} \Box$

\begin{theorem}[\cite{XY17}]If running the algorithm \textbf{FindStruct} on $F=(F_1,F_2,\cdots,F_n)\in C_{k,n}$ returns a set $A$, then for any $(a,i_1,i_2,\cdots,i_n)\in A$, any $0<\epsilon<1$, it holds that
\begin{equation*}
{\Large \rm Pr}\left[\frac{|\{x\in \mathbb{F}_2^k|F(x\oplus a)\oplus F(x)=i_1\cdots i_n\}|}{2^k}>1-n\epsilon\right]>\big(1-e^{-2p(n){\epsilon}^2}\big)^n.
\end{equation*}
Moreover, for any $j\in\{1,2,\cdots,n\}$, any $i_j\in\{0,1\}$ and any vector $a\in A_j^{i_j}$, it holds that
\begin{equation*}
{\Large \rm Pr}\left[\frac{|\{x\in \mathbb{F}_2^k|F_j(x\oplus a)\oplus F_j(x)=i_j\}|}{2^k}>1-\epsilon\right]>1-e^{-2p(n){\epsilon}^2}.
\end{equation*}
\end{theorem}

Before stating Theorem 3, we need to define a parameter. For any function $f\in C_{k,1}$, let
\begin{align}
\delta_f=\frac{1}{2^k} \max_{\substack{a\in \mathbb{F}_2^k\\ a\notin U_f}}\max_{i\in \mathbb{F}_2}|\{x\in \mathbb{F}_2^k|f(x\oplus a)+f(x)=i\}|.
\end{align}
For any function $F=(F_1,F_2,\cdots,F_n)\in C_{k,n}$, we define $\delta_F=\max_j\delta_{F_j}$, where $\delta_{F_j}$ is defined as Eq.(1). Obviously $\delta_F<1$. The larger $\delta_F$ is, the more difficult for excluding the vectors that are not linear structure of $F$ when applying the algorithm \textbf{FindStruct} on $F$.

\begin{theorem}[\cite{XY17}]Suppose $F\in C_{k,n}$ and $\delta_{F}\leq p_0<1$ for some constant $p_0$. If running the algorithm \textbf{FindStruct} on $F$ returns a set $A$, then for any $(a,i_1,i_2,\cdots,i_n)\in A$, it holds that
$$
{\rm Pr}\left[\,a\in U_F^{(i_1,\cdots,i_n)}\right]\geq 1-p_0^{p(n)}.
$$
That is, except for a probability of $p_0^{p(n)}$, the vectors in $A$ must be the linear structures of $F$.
\end{theorem}

Theorem 1 indicates that all linear structures of $F$ must be in the output set $A$. Noting that the vector $\bm{0}$ is a trivial linear structure of $F$, the set $A$ is always nonempty. Theorem 2 states that every vector in $A$ has a high probability of being an approximate linear structure of $F$. Theorem 3 shows that, except for a negligible probability, the vectors in the set $A$ output by the algoithm \textbf{FindStruct} with $p(n)=O(n)$ must be linear structures of $F$, under the condition that $\delta_{F}\leq p_0<1$ for some constant $p_0$.

By regarding each $F_j$ itself as a vector function that has only one component and applying Theorem 3 to $F_j$, we have following corollary:
\begin{corollary}
Suppose $F=(F_1,F_2,\cdots,F_n)\in C_{k,n}$ and $\delta_{F}\leq p_0<1$ for some constant $p_0$. The sets $A_j$ ($j=1,2,\cdots,n$), generated during running the algorithm \textbf{FindStruct} on $F$, satisfy that for any $a\in A_j$,
$$
{\rm Pr}\left[\,a\in U_{F_j}\right]\geq 1-p_0^{p(n)}.
$$
That is, except for a probability of $p_0^{p(n)}$, the vectors in $A_j$ must be the linear structures of $F_j$.
\end{corollary}
\section{Attack strategy}
In this section, we present a strategy for attacking general block ciphers using BV algorithm in the context of quantum related-key attack. We first describe the attack, then analyze under what conditions the attack will work and corresponding complexity of it.

\subsection{Description of the attack}
A general way to attack a symmetric cryptosystem $E$ using BV algorithm includes the following two steps:
\vskip 0.1cm

\noindent
\hangafter 1
\hangindent 1.4em
1. Construct a new function $F$ based on the cipher $E$ so that $F$ satisfies two conditions: (I) the attacker has quantum oracle access to $F$; (II) $F$ has a nontrivial linear structure that reveals the information of the secret key. Sometimes the linear structure itself is just the secret key.
\vskip 0.1cm

\noindent
\hangafter 1
\hangindent 1.4em
2. Apply the algorithm \textbf{FindStruct} to obtain the linear structure of $F$, and use it to recover the secret key.
\vskip 0.1cm

We now confine to the Electronic Codebook mode and give a specific attack strategy for block ciphers. Suppose $E_s:\mathbb{F}_2^n\rightarrow\mathbb{F}_2^n$ is a block cipher with a secret key $s\in\mathbb{F}_2^k$. Let $m$ be an arbitrary plaintext in the plaintext space. Define the function
\begin{align}
F_s^m:\mathbb{F}_2^k&\longrightarrow\mathbb{F}_2^n\\
x&\longrightarrow E_{x}(m)\oplus E_{s\oplus x}(m).\notag
\end{align}
Then for any $x\in\mathbb{F}_2^k$, we have $F_s^m(x\oplus s)\oplus F_s^m(x)=\bm{0}$. Therefore, the key $s$ is a nonzero linear structure of $F_s^m$. More precisely, $s\in U^{\bm{0}}_{F_s^m}$. Thus, we can find $s$ by applying the algorithm \textbf{FindStruct} to $F_s^m$. Since we have already know that $s$ is in $U^{\bm{0}}_{F_s^m}$, when running \textbf{FindStruct}, we only need to solve the linear equation group $\{x\cdot\omega=i_j|\omega\in W_j\}$ for $i_j=0$ in Step 2. The attack algorithm based on the simplified \textbf{FindStruct} algorithm is as follows:
\vskip 0.1cm

\begin{framed}

\noindent
Algorithm \textbf{RecoverKey}
\vskip 0.2cm

\noindent
\hangafter 1
\hangindent 1.4em
1. Choose a polynomial $p(n)$ and an arbitrary plaintext $m$. Define the function $F_s^m$ as Eq.(2). Denote $F_s^m=(F_{s,1}^{m},F_{s,2}^{m},\cdots,F_{s,n}^{m})$.

\noindent
\hangafter 1
\hangindent 1.4em
2. For $j=1,2,\cdots,n$, run BV algorithm on $F_{s,j}^{m}$ for $p(n)$ times to obtain a subset $W_j$ of $N_{F_{s,j}^{m}}$. The size of $W_j$ is $p(n)$.

\noindent
\hangafter 1
\hangindent 1.4em
3. For $j=1,2,\cdots,n$, solve the linear equation group $\{x\cdot\omega=0|\omega\in W_j\}$ to get the solution $A^0_j$.

\noindent
\hangafter 1
\hangindent 1.4em
4. Find the intersection $A=A_1^0\cap A_2^0\cap\cdots\cap A_n^0$. Verify the vectors in $A$ to determine the correct key.
\end{framed}
\vskip 0.1cm

The algorithm \textbf{RecoverKey} requires the quantum oracle access of $F_s^m$. The attacker can obtain this oracle by first querying the oracle $\mathcal{O}_{\mathcal{E}}$ to compute $|x,m,y\rangle\rightarrow|x,m,y\oplus E_{s\oplus x}(m)\rangle$, then implementing the unitary operator $U_E:|x,m,y\rangle\rightarrow|x,m,y\oplus E_{x}(m)\rangle$ by himself. The quantum circuit to implement $F_s^m$ is presented in Fig.5. Note that \textbf{RecoverKey} actually requires the quantum oracle access of $F_{s,j}^{m}$ for each $j\in\{1,2,\cdots,n\}$. Since we have assumed the attacker can query $\mathcal{O}_{\mathcal{E}_j}$ that returns solely $j$-${th}$ bit of $\mathcal{O}_{\mathcal{E}}$, this requirement can be satisfied.
\begin{figure}[H]
  \centering
  \includegraphics[width=15cm]{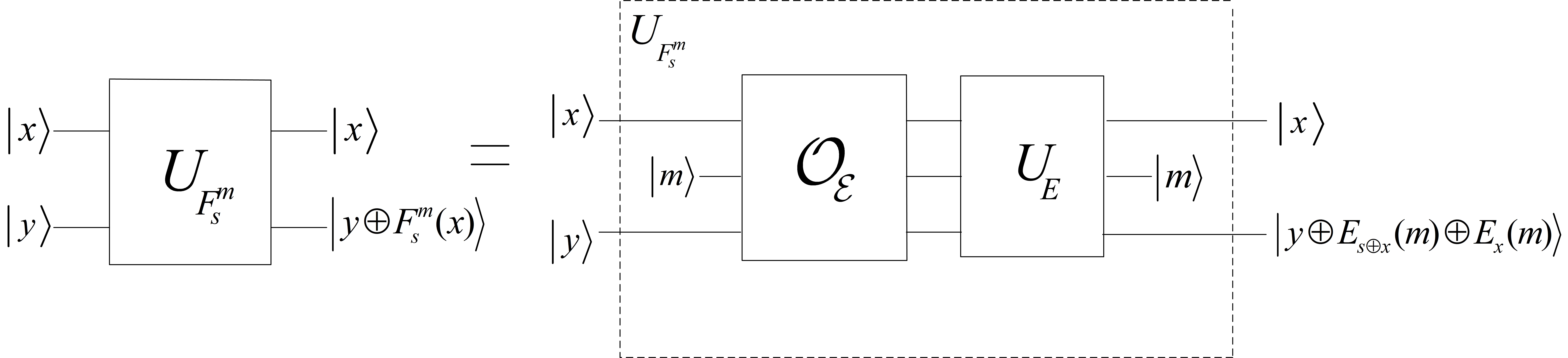}\\
  \caption{Quantum circuit to implement $F_s^m$}
\end{figure}

\subsection{Analysis of the attack}

We now analyze the performance of the algorithm \textbf{RecoverKey}, including the conditions under which the attack will work and the complexity. We first consider the case where \textbf{RecoverKey} is applied to a general block cipher $E$ without any restrictions. According to Theorem 1, the secret key $s$, as a linear structure of $F_s^m$, must be in the set $A=A_1^0\cap A_2^0\cap\cdots\cap A_n^0$. Therefore, by verifying all vectors in the set $A$, the attacker must be able to find the target key $s$. However, since there is no restriction on the block cipher $E$, the complexity of the algorithm \textbf{RecoverKey} may exceed the attacker's computational power.

To accurately compute the complexity of the algorithm \textbf{RecoverKey}, we separate it into three parts:

(1) executing BV algorithm for $np(n)$ times;

(2) solving $n$ linear equation groups;

(3) finding the intersection of $A_1^0$, $A_2^0$ $\cdots$ and $A_n^0$.

For the first part, running BV algorithm once needs to execute $2k+1$ Hadamard gates, one unitary operator $U_E$ and one quantum query on $\mathcal{O}_{\mathcal{E}}$. Thus, a total of $(2k+1+|E|_Q)np(n)$ universal gates and $np(n)$ quantum queries are needed. We assume a query requires one unit of time, then the complexity of this part is $O\big((2k+2+|E|_Q)np(n)\big)$. For the second part, the attacker needs to solve $n$ linear equation groups, and each one has $k$ variables and $p(n)$ equations. Solving a linear equation group with $k$ variables and $p(n)$ equations via Gaussian elimination method needs $O(p(n)k^2)$ calculations. Thus, the complexity of this part is $O(p(n)nk^2)$, which is a polynomial of $k$ and $n$. For the third part, the attacker needs to compute the intersection $A_1^0\cap A_2^0\cap\cdots\cap A_n^0$. Let $t=\max_j|A_j^0|$. Finding the intersection of these $n$ sets using sort method requires $O(nt\log t)$ calculations. The value of $t$ relies on the properties of $F_s^m$ and the value of $p(n)$. Since $A_j^0$ is the solution of a linear system with $p(n)$ equations, the size of $A_j^0$ should decrease rapidly as $p(n)$ increases. The larger $p(n)$ the attacker chooses, the smaller $t$ will be, so the attacker can choose a larger $p(n)$ to reduce $t$. (Even though this will increase the amount of unitary gates and queries required in the other two parts, the complexity of these two parts is still polynomial as long as $p(n)$ is still a polynomial.)

To sum up, the complexity of \textbf{RecoverKey} is $O\big((k^2+2k+2+|E|_Q)np(n)+nt\log t\big)$. It may be possible to choose a large $p(n)$ so that the parameter $t$ is a polynomial, but in the most general case we cannot guarantee that the algorithm \textbf{RecoverKey} can be efficiently executed.

Since we cannot bound the computational complexity of the algorithm \textbf{RecoverKey} when it is applied to a general block cipher, we consider the block ciphers with some restrictions. Specifically, we give two conditions. As long as $F_s^m$ satisfies one of them, then the algorithm \textbf{RecoverKey} can be executed efficiently with a high probability.
\vskip 0.2cm

\noindent
\textbf{$\bullet$ Condition 1: $\bm{\delta_{F_s^m}\leq p_0<1}$ for some constant $\bm{p_0}$.}
\vskip 0.2cm

Suppose $\delta_{F_s^m}\leq p_0<1$ for some constant $p_0$. By Corollary 1, if running the algorithm \textbf{RecoverKey} on $F_s^m$ with $p(n)=O(n)$, the set $A^0_j$, except for a negligible probability, will only contain the linear structures of $F_{s,j}^{m}$. In this situation, the complexity of \textbf{RecoverKey} is $O\big((k^2+2k+2+|E|_Q)n^2+nt\log t\big)$ and the value of $t=\max_j|A_j^0|$ is small. So as long as Condition 1 holds, the attack is valid and efficient with a high probability.

Condition 1 is a little abstract. To understand its cryptographic meaning, we compute the parameter $\delta_{F_s^m}$.
$$
\delta_{F_s^m}=\frac{1}{2^k}\max_j\max_{\substack{a\in \mathbb{F}_2^k\\ a\notin U_{F_{s,j}^{m}}}}\max_{i\in\mathbb{F}_2}|\{x\in\mathbb{F}_2^k|F_{s,j}^{m}(x)\oplus F_{s,j}^{m}(x\oplus a)=i\}|.
$$
Thus, Condition 1 means that there exist a constant $p_0$ such that, for any $j\in\{1,2,\cdots,n\}$, any $a\notin U_{F_{s,j}^{m}}$ and any $i\in\{0,1\}$, it holds that
\begin{equation}
\frac{|\{x\in\mathbb{F}_2^k|F_{s,j}^{m}(x)\oplus F_{s,j}^{m}(x\oplus a)=i\}|}{2^k}\leq p_0.
\end{equation}
Since $F_s^m(x)=E_{x}(m)\oplus E_{x\oplus s}(m)$, Eq.(3) is equivalent to
\begin{equation}
\frac{|\{x\in\mathbb{F}_2^k|E_{x,j}(m)\oplus E_{x\oplus s,j}(m)\oplus E_{x\oplus a,j}(m)\oplus E_{x\oplus s\oplus a}(m)=i\}|}{2^k}\leq p_0,
\end{equation}
where $E_{x,j}$ is the $j$-th component of $E_x$. Because $a\notin U_{F_{s,j}^{m}}$, we have that $a\neq\bm{0}$ and $a\neq s$, so $x$, $x\oplus s$, $x\oplus a$ and $x\oplus s\oplus a$ are always four different keys. Eq.(4) means that when averaging over all possible values of $x$, the exclusive value of the ciphertexts of $m$ under these four keys is not too biased. Generally speaking, a well constructed block cipher will not have obvious linearity, so Condition 1 is likely to be satisfied.
\vskip 0.2cm

\noindent
\textbf{$\bullet$ Condition 2: For each $\bm{j\in\{1,2,\cdots,n\}}$, $\bm{F_{s,j}^{m}}$ does not have many approximate linear structures.}
\vskip 0.2cm

More formally, Condition 2 requires that: there exists a sufficiently large polynomial $l(n)$ such that the amount of $\frac{1}{l(n)}$-close linear structures of each $F_{s,j}^{m}$ is small (at least smaller than some polynomial of $n$). Suppose Condition 2 holds. There exists such a polynomial $l(n)$. According to Theorem 2, any vector $a$ in the set $A_j^0$, which is generated during the execution of algorithm $\textbf{RecoverKey}$, satisfies that
$$
{\rm Pr}\left[\frac{|\{x\in\mathbb{F}_2^k|F_{s,j}^{m}(x\oplus a)\oplus F_{s,j}^{m}(x)=0\}|}{2^k}>1-\epsilon\right]>1-e^{-2p(n)\epsilon^2}.
$$
Let $\epsilon=\frac{1}{l(n)}$, $p(n)=nl(n)^2$, we have
$$
{\rm Pr}\left[\frac{|\{x\in\mathbb{F}_2^k|F_{s,j}^{m}(x\oplus a)\oplus F_{s,j}^{m}(x)=0\}|}{2^k}>1-\frac{1}{l(n)}\right]>1-e^{-2n}.
$$
That is, except for a negligible probability, $a$ is a $\frac{1}{l(n)}$-close linear structure of $F_{s,j}^{m}$. Therefore, if we run the algorithm $\textbf{RecoverKey}$ with $p(n)=nl(n)^2$, then except for a negligible probability, the size of the set $A^0_j$ will not be greater than the amount of $\frac{1}{l(n)}$-close linear structures of $F_{s,j}^{m}$. According to Condition 2, this amount is small, so the value of the parameter $t=\max_j|A_j^0|$ is small. Therefore, if Condition 2 holds, the complexity of \textbf{RecoverKey} with $p(n)=nl(n)^2$ is $O\big((k^2+2k+2+|E|_Q)n^2l(n)^2+nt\log t\big)$, where the parameter $t$ is a small number. This demonstrates that the algorithm \textbf{RecoverKey} is valid and efficient under Condition 2.

In fact, we can also analyze Condition 2 from the perspective of differential. If a vector $a$ is a $\frac{1}{l(n)}$-close linear structure of $F_{s,j}^{m}$, then $(a,0)$ is a differential of $F_{s,j}^{m}$ whose differential probability is greater than $1-\frac{1}{l(n)}$. If Condition 2 does not hold, it means $F_{s,j}^{m}$ has many high-probability differentials. $F_{s,j}^{m}(x)=E_{x\oplus s,j}(m)\oplus E_{x,j}(m)$. We can treat $F_{s,j}^{m}$ as a new cipher, and make chosen-plaintext query to $F_{s,j}^{m}$ by making related-key query to the original block cipher $E$. Thus, one can attack the original cipher $E$ by using $(a,0)$ to attack $F_{s,j}^{m}$. Based on above analysis, for a general well-constructed block cipher $E$, the amount of $\frac{1}{l(n)}$-close linear structures of each $F_{s,j}^{m}$ should be small, so Condition 2 is a reasonable requirement.
\vskip 0.2cm

The above two conditions are mild and should be satisfied by an well constructed block cipher. As long as the block cipher satisfies one of them, the algorithm \textbf{RecoverKey} can efficiently recover its key with a high probability.

\section{Discussion}

There remains many directions worth further studying. For example, there may exist other ways to construct a function that has a linear structure associated with the key based on the block cipher $E$. For the function $F_s^m$ constructed in this paper, the key $s$ is actually a special linear structure, i.e. a period. It may be possible to construct other functions that have a more general linear structure. For instance, consider the case $k=n$, namely, the case where the key length is equal to the blocksize. We can define the function $G_s^m(x)=E_{x\oplus s}(m)\oplus E_{x}(m)\oplus x$. Then for each $x\in\mathbb{F}_2^n$, we have $G_s^m(x\oplus s)\oplus G_s^m(x)=s$. Therefore, $s$ is a linear structure of $G_s^m$ and $s\in U_{G_s^m}^s$. Follow the usual attack strategy, the attacker can run the algorithm \textbf{FindStruct} on $G_s^m$. The vector $(s,s)$ must be in the output set $A$. Moreover, when the attacker compute the set $A_j=A_j^0\cup A_j^1$ in the step 2 of \textbf{FindStruct}, for any $a\in A_j^{i_j}$, if the $j$-th bit of $a$ is not equal to $i_j$, the attacker can discard it directly. This helps determine the target key $s$ faster. Whether there exists a construction of the function that can be proved to be optimal is also an interesting question. In addition, how to apply the algorithms proposed in this paper to specific practical block cipher worth investigating, too.

\section{Conclusion}
We apply Bernstein-Vazirani algorithm to related-key attack and propose a quantum attack for recovering the key of general block ciphers. We analyze under what conditions the attack will work, and rigorously compute its computational complexity. Our works show the power of relate-key attack in the quantum setting, and provides guidance for designing quantum-secure block ciphers.

\section*{Acknowledgement}
This work was supported by National Natural Science Foundation of China (Grant No.61672517), National Cryptography Development Fund (Grant No. MMJJ201 70108) and the Fundamental theory and cutting edge technology Research Program of Institute of Information Engineering, CAS (Grant No. Y7Z0301103).

\end{document}